\documentstyle[multicol,aps,psfig]{revtex}

\begin{document}

\title{Discovery of Jet Quenching and Beyond}
\author{Xin-Nian Wang} 
\address{
Nuclear Science Division, MS 70R0319,\\
Lawrence Berkeley National Laboratory, Berkeley, CA 94720}

%\date{\today}
\maketitle

\begin{abstract}

Recent observation of high-$p_T$ hadron spectra 
suppression and mono-jet production in central $Au+Au$ collisions
and their absence in $d+Au$ collisions at RHIC have confirmed the 
long predicted phenomenon of jet quenching in high-energy heavy-ion
collisions. Detailed analyses of the experimental data also show 
parton energy loss as the mechanism for the discovered jet quenching. 
Using a pQCD parton model that incorporates medium modified parton 
fragmentation functions and comparing to experimental data from 
deeply inelastic scattering off nuclei, one can conclude that the 
initial gluon (energy) density of the hot matter produced in 
central $Au+Au$ collisions that causes jet quenching at RHIC is 
about 30 (100) times higher than in a cold $Au$ nucleus. Combined 
with data on bulk and collective properties of the hot matter, 
the observed jet quenching provides strong evidence for the 
formation of a strongly interacting quark gluon plasma in 
central $Au+Au$ collisions at RHIC.

\end{abstract}

%\pacs{74.25.Fy, 72.15.Jf, 75.30.Kz}

% PACS, the Physics and Astronomy Classification Scheme.

\begin{multicols}{2}

\section{Introduction}

In the last three years, since the physics operation of the Relativistic 
Heavy-ion Collider (RHIC) at Brookhaven National Laboratory (BNL), one 
has witnessed tremendous progress in heavy-ion experimental physics and
many milestones in the search for the elusive quark gluon plasma (QGP) 
that is expected to be formed in high-energy heavy-ion collisions. One 
of the most exciting phenomena observed at RHIC is jet quenching, or
suppression of leading hadrons from fragmentation of hard partons
due to their strong interaction with the dense medium. Such a phenomenon
was long predicted by a pQCD-based model calculation \cite{wg92}, but
was not seen in high-energy heavy-ion collisions until the first
measurements in central $Au+Au$ collisions at $\sqrt{s}=130$ GeV at 
RHIC \cite{phenix-r1,star-r1}. The azimuthal distribution of 
high $p_T$ hadrons was also found to display large anisotropy 
with respect to the reaction planes \cite{star-jetv2}
of non-central $Au+Au$ collisions which was expected to be caused 
by the path-length dependence of jet quenching \cite{wangv2,gvwv2}.

The measurements were confirmed and further extended to a larger $p_T$
range in $Au+Au$ collisions at $\sqrt{s}=200$ GeV, and the
suppression was found to have a weak $p_T$-dependence at
$p_T>6$ GeV/$c$ \cite{phenix-r2,star-r2}, independent of hadron species.
At the same time, the back-side high-$p_T$ two-hadron correlation
in azimuthal angle, characteristic of high-$p_T$ back-to-back jets 
in $p+p$ collisions, was found to disappear in central $Au+Au$ 
collisions \cite{star-jet}, confirming the predicted mono-jet phenomenon
of jet quenching \cite{monojet}. 

The final milestone in the experimental discovery of jet 
quenching was achieved during the third year of the RHIC
physics program, when both the single-hadron 
spectra \cite{dauphenix,daustar,dauphobos} and the disappearance 
of away-side two-hadron correlation \cite{daustar} were found to be
absent in the same central rapidity region of $d+Au$ collisions 
at $\sqrt{s}=200$ GeV. These $d+Au$ results prove that the 
observed high-$p_T$ suppression patterns in $Au+Au$ 
collisions are not initial state effects encoded in the 
wavefunction of a beam nucleus, but are jet quenching caused by 
final state interaction of hard partons with the produced dense 
medium.

Following these three major experimental results, additional
data{\bf ---}such as the dependence of away-side suppression on
the azimuthal angle relative to the reaction plane \cite{filimonov},
modification of the hadron distributions (fragmentation functions)
both along and opposite the direction of the triggered high-$p_T$ 
hadron \cite{fqwang}, and absence of suppression in the direct
photon spectra \cite{frantz} in $Au+Au$ collisions{\bf ---}have now
solidified the conclusion that the observed jet quenching is caused by 
parton multiple scattering and energy loss in the hot and dense medium.
Furthermore, the splitting of baryon and meson spectra suppression
and azimuthal anisotropy in the intermediate $p_T<6$ GeV/$c$ 
region \cite{phenix-fl,phenix-fl2,star-fl} also point to
non-trivial medium modification of hadronization that is
indicative of dense partonic matter. Such a wealth of data affords 
one a quantitative phenomenological analysis of jet 
quenching and a tomographical picture of the hot and dense matter
formed in heavy-ion collisions at RHIC.

In this report, we will first briefly review the recent progress in 
pQCD study of parton multiple scattering and induced radiative 
energy loss in a dense medium. Combining this with the pQCD parton
model of high-$p_T$ jet and hadron production, one can analyze 
the observed jet quenching phenomena to extract properties of the
dense matter produced.
One can combine the deduced properties with other information 
from analyses of bulk particle spectra such as collective 
flow and total multiplicity and energy production to present
a  collection of compelling evidence for the existence of a 
strongly interacting quark gluon plasma in $Au+Au$ collisions
at RHIC. However, we will focus in this report only on the analysis of 
jet quenching data and the properties of the dense matter one
can extract from these data.

\section{Parton Energy Loss}

One important step in establishing evidence of QGP formation is to
characterize the properties of the produced dense medium, for example, 
the parton and energy density and current-current correlations, 
among many other characteristics. Traditionally, one can study the 
properties of a medium via scattering experiments with particles. 
In deeply inelastic scattering (DIS) experiments, for example, 
leptons scatter off the nucleon medium via photon exchange. 
The response function or the correlation function of the electromagnetic 
currents $j^{em}_\mu(x)=\sum_q e_q\bar{\psi}_q(x)\gamma_\mu\psi_q(x)$,
\begin{equation}
W_{\mu\nu}(q)=\frac{1}{4\pi}\int d^4x e^{iq\cdot x}\langle A\mid j^{em}_\mu(0)
j^{em}_\nu(x)\mid A \rangle ,
\end{equation}
is a direct measurement of the quark distributions in a nucleon or nucleus.
For a dynamic system in heavy-ion collisions, one can no longer 
use the technique of scattering with an external beam of particles in
the conventional sense because of the transient nature of the matter.
The lifetime of the system is very short, on the order of a 
few fm/$c$. The initial spatial size is only the size of the 
colliding nuclei, about 6 fm in diameter in the transverse
dimension for the heaviest nuclei. The system expands very rapidly both 
in the longitudinal and the transverse direction. These characteristics 
make it impossible to use external probes to study the properties 
of the produced dense matter in high-energy heavy-ion
collisions. One has to resort to internal probes
such as the electromagnetic emission from the dynamic medium, 
whose rate is given by the thermal average of the above
correlation function. The emission rate depends mainly on the 
local temperature or the parton density while the total yield 
also depends on the whole evolution history of the system. 
Therefore, a strongly interacting system can reveal
its properties and dynamics through photon and dilepton emission.
The same medium interaction should also cause attenuation of 
fast and energetic partons propagating through the medium. 
Such an effect is the underlying physics of the jet 
quenching \cite{bj,Gyulassy:1990ye} phenomenon and jet 
tomography technique for studying properties of dense matter 
in high-energy heavy-ion collisions.

Jet quenching as a probe of the dense matter in heavy-ion collisions
takes advantage of the hard processes of jet production in high-energy 
heavy-ion collisions. Because large-$p_T$ partons are produced
very early in heavy-ion collisions, they can probe the early
stage of the formed dense medium. Similar to the technology of 
computed tomography (CT), the study of these energetic particles,
their initial production and interaction with the dense medium can yield
critical information about the properties of the medium that is otherwise
difficult to access through soft hadrons from the hadronization of the
bulk matter. Though relatively rare with small cross
sections, the jet production rate can be calculated perturbatively 
in QCD and agrees well with experimental measurements in high-energy
$p+p(\bar{p})$ collisions. A critical component of the jet tomography is
then to understand the jet's attenuation through dense matter as it
propagates through the medium.

There have been many theoretical 
studies \cite{Baier:2000mf,Gyulassy:2003mc,Kovner:2003zj} of jet 
attenuation in a hot medium in recent years. The first attempt was by 
Bjorken \cite{bj} to calculate elastic energy
loss of a parton via elastic scattering in the hot medium. A simple
estimate can be given by the thermal averaged energy transfer
$\nu_{\rm el}\approx q_\perp^2/2\omega$ of the jet parton to a thermal
parton, with energy $\omega$, and $q_\perp$ the transverse momentum
transfer of the elastic scattering. The resultant elastic energy 
loss \cite{wangrep}
\begin{equation}
\frac{dE_{\rm el}}{dx}=C_2\frac{3\pi\alpha_{\rm s}^2}{2}T^2
\ln\left(\frac{3ET}{2\mu^2}\right)
\end{equation}
is sensitive to the temperature of the thermal medium but is 
small compared to radiative energy loss. Here, $\mu$ is the Debye screening
mass and $C_2$ is the Casimir coefficient of the propagating 
parton in its fundamental presentation. The elastic energy loss 
can also be calculated within finite temperature field theory of
QCD \cite{thoma} with a more careful and consistent treatment of 
the Debye screening effect.

Though there had been estimates of the radiative parton energy loss 
using the uncertainty principle \cite{bh}, the first theoretical study 
of QCD radiative parton energy loss incorporating
Landau-Pomeranchuk-Migdal interference effect \cite{lpm} is by Gyulassy and
Wang \cite{gw94} where multiple parton scattering is modeled by a screened
Coulomb potential model. Baier {\it et al.} (BDMPS) \cite{bdmps} later 
considered the effect of gluon rescattering which turned out to be very 
important for gluon radiation induced by multiple scattering in a dense
medium. These two studies have ushered in many recent works on the subject,
including a path integral approach to the problem \cite{zakharov},
twist \cite{wgdis} or opacity expansion framework \cite{glv,wied} 
which is more suitable for multiple parton scattering in a finite
system. The radiative parton energy loss in the leading order of the 
twist or opacity expansion was found to have a simple 
form \cite{glv}
\begin{equation}
\frac{dE_{\rm rad}}{dx}\approx C_2 \frac{\alpha_{\rm s} \mu^2}{4}
\frac{L}{\lambda} \ln\left(\frac{2E}{\mu^2 L}\right),
\label{eloss0}
\end{equation}
for an energetic parton ($E\gg \mu$) in a static medium, 
where $\lambda$ is the gluon's mean free path in the medium. The 
unique $L$-dependence of the parton energy loss, which was first 
discovered by BDMPS \cite{bdmps} is a consequence of the non-Abelian 
LMP interference effect in a QCD medium. In a dynamic system the
total energy loss was found \cite{gvwv2,ww02,salgado} to be 
proportional to a line integral of the gluon density along the 
parton propagation path.

Since gluons are bosons, there should also be stimulated gluon 
emission and absorption by the propagating parton in 
the presence of thermal gluons in the hot medium.  Such detailed
balance is crucial for parton thermalization and should also be 
important for calculating the energy loss of an energetic parton 
in a hot medium. Taking into account such detailed balance in
gluon emission, Wang and Wang \cite{ww01} obtained an asymptotic 
behavior of the effective energy loss in the opacity expansion 
framework ,
\begin{eqnarray}
   {\Delta E\over E}\approx &&
   {{\alpha_s C_2 \mu^2 L^2}\over 4\lambda E}
   \left[\ln{2E\over \mu^2L} -0.048\right]\nonumber \\
   &&\hspace{-0.3in}-
   {{\pi\alpha_s C_F}\over 3} {{LT^2}\over {\lambda E^2}}
   \left[
   \ln{{\mu^2L}\over T} -1+\gamma_{\rm E}-{{6\zeta^\prime(2)}\over\pi^2}
\right],
 \end{eqnarray}
where the first term is from the induced bremsstrahlung and the second
is due to gluon absorption in detailed balance, which effectively
reduces the total parton energy loss in the medium. Though the effect
of detailed balance is small at large energy ($E\gg \mu$), it
changes the effective energy dependence of the energy loss
in the intermediate energy region.

\section{Modified Fragmentation Function}

Because of color confinement in the vacuum, one can never
separate hadrons fragmenting from the leading parton and
particles materializing from the radiated gluons. The total
energy in the conventionally defined jet cone in principle 
should not change due to induced radiation, assuming that most of 
the energy carried by radiative gluons remains inside the 
jet cone \cite{baier99}. Additional rescattering of the
emitted gluon with the medium could broaden the jet cone
significantly, thus reducing the energy in a fixed cone. 
However, fluctuation of the underlying
background in high-energy heavy-ion collisions makes it very 
difficult, if not impossible, to determine the energy of a jet 
on an event-by-event base with sufficient precision to discern a finite
energy loss of the order of 10 GeV. Since high-$p_T$ hadrons
in hadronic and nuclear collisions come from fragmentation of 
high-$p_T$ jets, energy loss naturally leads to suppression 
of high-$p_T$ hadron spectra.  This was why Gyulassy and Wang 
proposed \cite{wg92} to  measure the suppression of high-$p_T$ 
hadrons to study parton energy loss in heavy-ion collisions.

Since parton energy loss effectively slows down the leading parton in
a jet, a direct manifestation of jet quenching is the modification of the 
jet fragmentation function, $D_{a\rightarrow h}(z,\mu^2)$, which can 
be measured directly in events in which one can identify the jet via a
companion particle like a direct photon \cite{wh} or a triggered high
$p_T$ hadron. This modification can be directly translated into the 
energy loss of the leading parton. Since inclusive hadron
spectra are a convolution of the jet production cross section and the
jet fragmentation function in pQCD, the suppression of 
inclusive high-$p_T$ hadron spectra is a direct consequence 
of the medium modification of the jet fragmentation function 
caused by parton energy loss.

Since a jet parton is always produced via a hard process involving a
large momentum scale, it should also have final state radiation with
and without rescattering, leading to the DGLAP evolution of
fragmentation functions. Such final state radiation effectively
acts as a self-quenching mechanism, softening the leading hadron
distribution. This process is quite similar to the induced 
gluon radiation and the two should have a strong interference
effect. It is therefore natural to study jet quenching
and modified fragmentation functions in the framework of modified
DGLAP evolution equations in a medium \cite{wgdis}.

To demonstrate medium modified fragmentation function and
parton energy loss, we review here the twist expansion approach 
in the study of deeply inelastic scattering (DIS) $eA$ \cite{wgdis,bwzxnw}. 
While the results can be readily extended to the case of  a hot
medium, they also provide a baseline for comparison to parton
energy loss in cold nuclei.

In the parton model with the collinear factorization approximation,
the leading-twist contribution to the semi-inclusive cross section
can be factorized into a product of parton distributions,
parton fragmentation functions and the hard partonic cross section.
Including all leading log radiative corrections, the lowest order
contribution from a single
hard $\gamma^*+ q$ scattering can be written as
\begin{eqnarray}
\frac{dW^S_{\mu\nu}}{dz_h}
&= &\sum_q e_q^2 \int dx f_q^A(x,\mu_I^2) H^{(0)}_{\mu\nu}(x,p,q)
 D_{q\rightarrow h}(z_h,\mu^2)\, , \nonumber
\end{eqnarray}
where $H^{(0)}_{\mu\nu}(x,p,q)$ is the hard part of the
process in leading order, $\ell_h$ the observed hadron 
momentum, $p = [p^+,0,{\bf 0}_\perp] \label{eq:frame}$
the momentum per nucleon in the nucleus,
$q = [-Q^2/2q^-, q^-, {\bf 0}_\perp]$ the momentum transfer 
carried by the virtual photon. In the chosen frame, $q^-$ is the
quark momentum transferred from the virtual photon. The momentum 
fraction carried by the hadron is defined as
$z_h=\ell_h^-/q^-$ and $x=x_B=Q^2/2p^+q^-$ is the Bjorken variable.
$\mu_I^2$ and $\mu^2$ are the factorization scales for the initial
quark distributions $f_q^A(x,\mu_I^2)$ in a nucleus and the fragmentation
functions $D_{q\rightarrow h}(z_h,\mu^2)$, respectively.

In a nuclear medium, the propagating quark in DIS will experience additional
scattering with other partons from the nucleus. The rescatterings may
induce additional gluon radiation and cause the leading quark to lose
energy. Such induced gluon radiations will effectively give rise to
additional terms in a DGLAP-like evolution equation leading to the 
modification of the fragmentation functions in a medium. These are 
the so-called higher-twist corrections since they involve higher-twist 
parton matrix elements and are power-suppressed. The leading contributions
involve two-parton correlations from two different nucleons inside
the nucleus.

One can apply the generalized factorization \cite{LQS} to 
these multiple scattering processes. In this approximation, the 
double scattering contribution to radiative correction can 
be calculated and the effective modified fragmentation 
function is \cite{wgdis}
\begin{eqnarray}
\widetilde{D}_{q\rightarrow h}(z_h,\mu^2)&\equiv&
D_{q\rightarrow h}(z_h,\mu^2)
+\int_0^{\mu^2} \frac{d\ell_T^2}{\ell_T^2}
\frac{\alpha_s}{2\pi} \int_{z_h}^1 \frac{dz}{z} \nonumber \\
& & \left[ \Delta\gamma_{q\rightarrow qg}(z,x,x_L,\ell_T^2)\right.
 D_{q\rightarrow h}(z_h/z)  \nonumber \\
&+& \left. \Delta\gamma_{q\rightarrow gq}(z,x,x_L,\ell_T^2)
D_{g\rightarrow h}(z_h/z)\right] \, , \label{eq:MDq}
\end{eqnarray}
in much the same form as the DGLAP correction in vacuum,
where $D_{q\rightarrow h}(z_h,\mu^2)$ and
$D_{g\rightarrow h}(z_h,\mu^2)$ are the leading-twist
quark and gluon fragmentation functions. The modified 
splitting functions are given as
\begin{eqnarray}
\Delta\gamma_{q\rightarrow qg}(z,x,x_L,\ell_T^2)&=&
\left[\frac{1+z^2}{(1-z)_+}T^{A}_{qg}(x,x_L) \right. \nonumber \\
& & \hspace{-0.8in}+\left. \delta(1-z)\Delta T^{A}_{qg}(x,\ell_T^2) \right]
\frac{2\pi\alpha_s C_A}
{\ell_T^2 N_c f_q^A(x,\mu_I^2)}\, ,
\label{eq:r1}\\
\Delta\gamma_{q\rightarrow gq}(z,x,x_L,\ell_T^2)
&=& \Delta\gamma_{q\rightarrow qg}(1-z,x,x_L,\ell_T^2). \label{eq:r2}
\end{eqnarray}
Here, the fractional momentum $x=x_B$ is carried by the initial 
quark and $x_L =\ell_T^2/2p^+q^-z(1-z)$ is the additional momentum
fraction carried either by the quark or gluon in the
secondary scattering that is required for induced gluon
radiation. The twist-four parton matrix element of the nucleus, 
\begin{eqnarray}
T^{A}_{qg}(x,x_L)&=& \int \frac{dy^{-}}{2\pi}\, dy_1^-dy_2^-
e^{i(x+x_L)p^+y^-}(1-e^{-ix_Lp^+y_2^-})\nonumber  \\
& & \hspace{-0.4in}\times (1-e^{-ix_Lp^+(y^--y_1^-)})
\theta(-y_2^-)\theta(y^- -y_1^-)\nonumber  \\
&&\hspace{-0.4in}\times\frac{1}{2}\langle A | \bar{\psi}_q(0)\,
\gamma^+\, F_{\sigma}^{\ +}(y_{2}^{-})\, F^{+\sigma}(y_1^{-})\,\psi_q(y^{-})
| A\rangle  \;\; , \label{Tqg}
\end{eqnarray}
has a dipole-like structure which is a result of LPM interference 
in gluon bremsstrahlung. 

\begin{figure}
\centerline{\psfig{figure=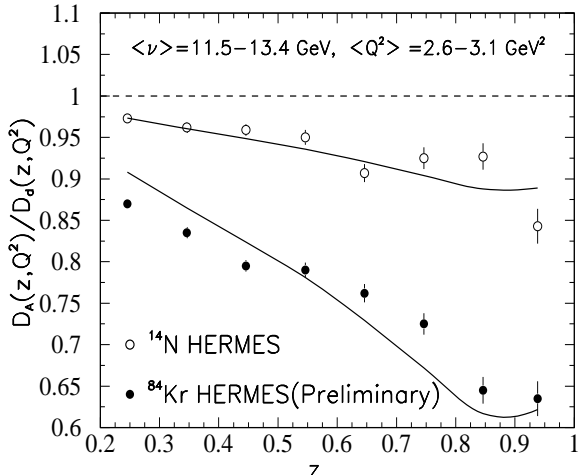,width=3.0in,height=2.5in}}
\baselineskip=10pt
\caption{Predicted nuclear modification of the jet fragmentation functions
compared to the HERMES data \protect\cite{hermes} on ratios of
hadron distributions between $A$ and $d$ targets in DIS.
\label{fig:hermes1}}
\end{figure}

Averaged over a Gaussian nuclear distribution, the interference 
will produce a factor $1-e^{-x_L^2/x_A^2}$ with $x_A=1/MR_A$. Here
$R_A$ is the nuclear size and $M$ is the nucleon mass.
Using the factorization approximation\cite{wgdis,LQS,ow}, one can
relate the twist-four parton matrix elements of the nucleus
to the twist-two parton distributions of nucleons and the nucleus,
\begin{equation}
T^{A}_{qg}(x,x_L)\approx \frac{\widetilde{C}}{x_A}
(1-e^{-x_L^2/x_A^2}) f_q^A(x),
\label{modT2}
\end{equation}
where $\widetilde{C}\equiv 2C x_Tf^N_g(x_T)$ is considered a constant.
One can identify $1/x_Lp^+=2q^-z(1-z)/\ell_T^2$ 
as the formation time of the emitted gluons. In the limit of collinear
radiation ($x_L\rightarrow 0$) or when the formation time of the
gluon radiation is much larger than the nuclear size, the above matrix 
element vanishes, demonstrating a typical LPM interference effect.

Since the LPM interference suppresses gluon radiation whose 
formation time ($\tau_f \sim Q^2/\ell_T^2p^+$) is larger than 
the nuclear size $MR_A/p^+$ (here in the infinite momentum frame), 
$\ell_T^2$ should then have a 
minimum value of $\ell_T^2\sim Q^2/MR_A\sim Q^2/A^{1/3}$. 
Therefore, the leading higher-twist contribution is  proportional 
to $\alpha_s R_A/\ell_T^2 \sim \alpha_s R_A^2/Q^2$
due to double scattering and depends quadratically 
on the nuclear size $R_A$.

\begin{figure}
\centerline{\psfig{figure=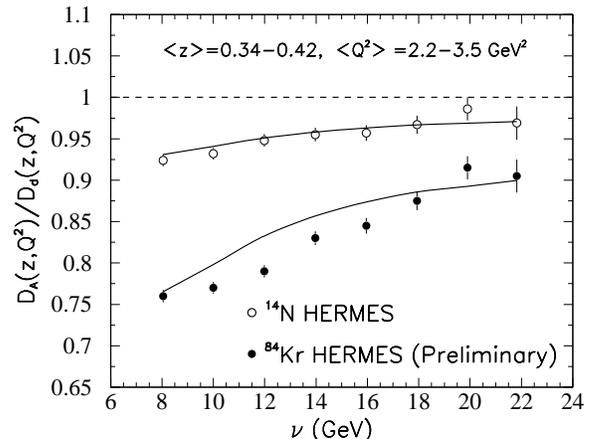,width=3.0in,height=2.3in}}
\caption{Energy dependence of the nuclear modification compared
with the HERMES data \protect\cite{hermes}.}
\label{fig:hermes2}
\end{figure}

With the assumption of the factorized form 
of the twist-4 nuclear parton matrices, there is only one free 
parameter $\widetilde{C}(Q^2)$
which represents quark-gluon correlation strength inside nuclei.
Once it is fixed, one can predict the $z$, energy and
nuclear dependence of the medium modification of the fragmentation
function. Shown in Fig.~\ref{fig:hermes1} are the calculated  \cite{ww02}
nuclear modification factor of the fragmentation functions for $^{14}N$ 
and $^{84}Kr$ targets as compared to the recent HERMES data\cite{hermes}.
The predicted shape of the $z$-dependence agrees well 
with the experimental data.  A remarkable feature of the prediction
is the quadratic $A^{2/3}$ nuclear size dependence, which is verified 
for the first time by an experiment.
By fitting the overall suppression for one nuclear target, 
one obtains the only parameter in the calculation,
$\widetilde{C}(Q^2)=0.0060$ GeV$^2$ 
with $\alpha_{\rm s}(Q^2)=0.33$ at $Q^2\approx 3$ GeV$^2$.
The predicted energy ($\nu$) dependence also agrees with 
the experimental data as shown in Fig.~\ref{fig:hermes2}. 
Note that the suppression goes away as the quark energy increases.
This is in sharp contrast to the jet quenching in heavy-ion collisions.

One can quantify the modification of the fragmentation
by the quark energy loss which is defined as the momentum fraction
carried by the radiated gluon,
\begin{eqnarray}
\langle\Delta z_g\rangle
&=& \int_0^{\mu^2}\frac{d\ell_T^2}{\ell_T^2}
\int_0^1 dz \frac{\alpha_s}{2\pi}
 z\,\Delta\gamma_{q\rightarrow gq}(z,x_B,x_L,\ell_T^2) \nonumber \\
&=&\widetilde{C}\frac{C_A\alpha_s^2}{N_c}
\frac{x_B}{x_AQ^2} \int_0^1 dz \frac{1+(1-z)^2}{z(1-z)}\nonumber \\
&\times&\int_0^{x_\mu} \frac{dx_L}{x_L^2}(1-e^{-x_L^2/x_A^2}),
\label{eq:heli-loss}
\end{eqnarray}
where $x_\mu=\mu^2/2p^+q^-z(1-z)=x_B/z(1-z)$ if the
factorization scale is set as $\mu^2=Q^2$. When $x_A\ll x_B\ll 1$ 
the leading quark energy loss is approximately
\begin{eqnarray}
\langle \Delta z_g\rangle(x_B,\mu^2)& \approx &
\widetilde{C}\frac{C_A\alpha_s^2}{N_c}\frac{x_B}{Q^2
x_A^2}6\sqrt{\pi}\ln\frac{1}{2x_B}\, .
\label{eq:appr1-loss}
\end{eqnarray}
Since $x_A=1/MR_A$, the energy loss $\langle \Delta
z_g\rangle$ thus depends quadratically on the nuclear size.

In the rest frame of the nucleus, $p^+=M$, $q^-=\nu$, and
$x_B\equiv Q^2/2p^+q^-=Q^2/2M\nu$. One can
get the averaged total energy loss as
$ \Delta E=\nu\langle\Delta z_g\rangle
\approx  \widetilde{C}(Q^2)\alpha_{\rm s}^2(Q^2)
MR_A^2(C_A/N_c) 3\ln(1/2x_B)$.
With the determined value of $\widetilde{C}$, 
$\langle x_B\rangle \approx 0.124$ in the HERMES experiment\cite{hermes}
and the average distance $\langle L_A\rangle=R_A\sqrt{2/\pi}$
for the assumed Gaussian nuclear distribution,
one gets the quark energy 
loss $dE/dL\approx 0.5$ GeV/fm inside a $Au$ nucleus.

\section{Jet Quenching in Hot Medium}

To extend the study of modified fragmentation functions to 
jets in heavy-ion collisions, one can
assume $\langle k_T^2\rangle\approx \mu^2$ (the Debye screening mass)
and a gluon density profile
$\rho(y)=(\tau_0/\tau)\theta(R_A-y)\rho_0$ for a 1-dimensional 
expanding system. Since the initial jet production 
rate is independent of the final gluon density, which can be 
related to the parton-gluon scattering cross 
section\cite{Baier:1996sk} 
[$\alpha_s x_TG(x_T)\sim \mu^2\sigma_g$], one has then
\begin{equation}
\frac{\alpha_s T_{qg}^A(x_B,x_L)}{f_q^A(x_B)} \sim
\mu^2\int dy \sigma _g \rho(y)
[1-\cos(y/\tau_f)],
\end{equation}
where $\tau_f=2Ez(1-z)/\ell_T^2$ is the gluon formation time.
The averaged fractional energy loss is then,
\begin{eqnarray}
\langle\Delta z_g\rangle &=&\frac{C_A\alpha_s}{\pi}
\int_0^1 dz \int_0^{Q^2/\mu^2}du \frac{1+(1-z)^2}{u(1+u)}\nonumber \\
&\times& \int_{\tau_0}^{R_A} d\tau\sigma_g\rho(\tau) 
\left[1-\cos\left(\frac{(\tau-\tau_0)\,u\,\mu^2}{2Ez(1-z)}\right)\right].
\end{eqnarray}
Keeping only the dominant contribution and assuming 
$\sigma_g\approx C_a 2\pi\alpha_s^2/\mu^2$ ($C_a$=1 for $qg$ and 9/4 for
$gg$ scattering), one obtains the total energy loss,
\begin{equation}
\langle \Delta E \rangle \approx \pi C_aC_A\alpha_s^3
\int_{\tau_0}^{R_A} d\tau \rho(\tau) (\tau-\tau_0)\ln\frac{2E}{\tau\mu^2}.
\label{effloss}
\end{equation}
which has also be obtained in the opacity expansion 
approach\cite{gvwv2,salgado} in a thin plasma.

In a static medium, the gluon density is independent of time and
one can recover the quadratic lenghth dependence in Eq.~(\ref{eloss0}).
Neglecting the logarithmic dependence on $\tau$, the averaged energy 
loss can be expressed as
\begin{equation}
\langle\frac{dE}{dL}\rangle_{1d} \approx (dE_0/dL) (2\tau_0/R_A)
\label{1d-loss}
\end{equation}
in a 1-dimensional expanding system, $\rho(\tau)=\rho_0\tau_0/\tau$.
Here $dE_0/dL\propto \rho_0R_A$ is the energy loss in a static 
medium with the same gluon density $\rho_0$ as in the 1-d expanding 
system at time $\tau_0$. Because of the expansion, the averaged 
energy loss $\langle dE/dL\rangle_{1d}$ is suppressed as compared 
to the static case and does not depend linearly on the system size
at a fixed value of initial gluon density.

One should emphasize now that the above simple logarithmic energy 
dependence of the parton energy loss in Eqs.~(\ref{eq:appr1-loss})
and (\ref{effloss}) is only an asymptotic behavior. However,
kinematic limits on induced gluon radiation from a parton with
finite energy can result in a stronger energy dependence \cite{glv}.
The effect of detailed balance with thermal absorption further
increases the energy dependence \cite{ww01}, while reducing the effective
parton energy loss for intermediate values of parton energy.
Such a detailed balance effect sets the energy dependence
of the energy loss in cold nuclei in DIS apart from the hot medium in
heavy-ion collisions. If one parameterizes the energy dependence 
of the energy loss including the full kinematic limits and thermal 
absorption, one would get
\begin{equation}
 \langle\frac{dE}{dL}\rangle_{1d}=\epsilon_0 (E/\mu-1.6)^{1.2}
 /(7.5+E/\mu).
\label{eq:loss}
\end{equation}
The threshold is the consequence of gluon absorption that competes
with radiation and effectively shuts off the energy loss. The
parameter $\mu$ is set to be 1.5 GeV in the following discussions.
Such a detailed balance effect can also explain why the total hadron
multiplicity that is dominated by soft hadrons does have
significant enhancement over the $p+p$ collisions as result of induced
gluon emission.

\subsection{Single Spectra}

To calculate the modified high-$p_T$ spectra in $A+A$ collisions,
we use a LO pQCD model \cite{Wang:1998ww},
\begin{eqnarray}
  \frac{d\sigma^h_{AA}}{dyd^2p_T}&=&K\sum_{abcd} 
  \int d^2b d^2r dx_a dx_b d^2k_{aT} d^2k_{bT}  \nonumber \\
  &\times& t_A(r)t_A(|{\bf b}-{\bf r}|) 
  g_A(k_{aT},r)  g_A(k_{bT},|{\bf b}-{\bf r}|) \nonumber \\
  &\times& f_{a/A}(x_a,Q^2,r)f_{b/A}(x_b,Q^2,|{\bf b}-{\bf r}|) \nonumber \\
 &\times& \frac{D_{h/c}^\prime (z_c,Q^2,\Delta E_c)}{\pi z_c}  
  \frac{d\sigma}{d\hat{t}}(ab\rightarrow cd), \label{eq:nch_AA}
\end{eqnarray}
with medium modified fragmentation functions $D_{h/c}^\prime$.
Here, $z_c=p_T/p_{Tc}$, $y=y_c$, $\sigma(ab\rightarrow cd)$ are 
elementary parton scattering cross sections and $t_A(b)$ is the 
nuclear thickness function normalized to $\int d^2b t_A(b)=A$.
The $K\approx 1.5$--2 factor is used to account for higher order pQCD 
corrections.

\begin{figure}
\centerline{\psfig{figure=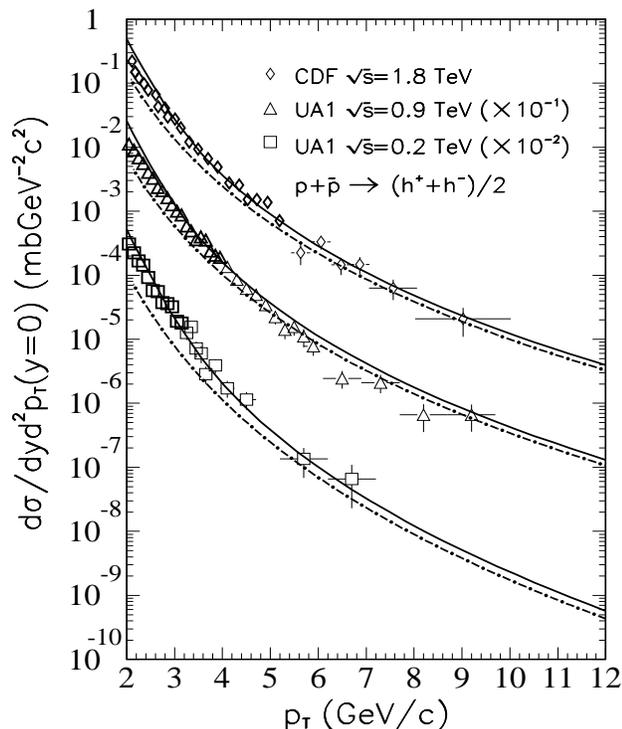,width=3.2 in,height=3.8in}}
\caption{ Single-inclusive spectra of charged hadrons in $p+\bar{p}$
  collisions at  $\sqrt{s}=200, 900, 1800$ GeV. The solid (dot-dashed) 
lines are parton model calculations with (without) intrinsic $k_T$. 
Experimental data are from Refs.~\protect\cite{Albajar:1989an,Abe:1988yu}.}
\label{fig-ua1th}
\end{figure}

To simplify the incorporation of the medium modification of the
fragmentation function in the parton model, an effective model \cite{wh} 
can be used:
\begin{eqnarray}
D_{h/c}^\prime(z_c,Q^2,\Delta E_c) 
&=&e^{-\langle\frac{\Delta L}{\lambda}\rangle}D^0_{h/c}(z_c,Q^2)
+(1-e^{-\langle \frac{\Delta L}{\lambda}\rangle})\nonumber \\
&& \hspace{-0.8in}
\times\left[ \frac{z_c^\prime}{z_c} D^0_{h/c}(z_c^\prime,Q^2) 
+\langle \frac{\Delta L}{\lambda}\rangle
\frac{z_g^\prime}{z_c} D^0_{h/g}(z_g^\prime,Q^2)\right]
\label{modfrag} 
\end{eqnarray}
where $z_c^\prime,z_g^\prime$ are the rescaled momentum fractions
after parton energy loss.
This effective model is found \cite{wang-bielefeld} to reproduce 
the pQCD result from Eq.(\ref{eq:MDq}) very well, but only when
$\Delta z=\Delta E_c/E$ is set to
 be $\Delta z\approx 0.6 \langle z_g\rangle$.
Therefore the actual averaged parton energy loss should be
$\Delta E/E=1.6\Delta z$, with $\Delta z$ extracted from the 
effective model. The factor 1.6 is mainly caused by the unitarity
correction effect in the pQCD calculation. The fragmentation 
functions in  free space $D^0_{h/c}(z_c,Q^2)$ will be given by 
the BBK  parameterization \cite{bkk}.

The parton distributions per nucleon $f_{a/A}(x_a,Q^2,r)$
inside the nucleus are assumed to be factorizable into the parton 
distributions in a free nucleon given by the MRSD$-^{\prime}$  
parameterization \cite{martin} and the impact-parameter dependent 
nuclear modification factor given by the new 
HIJING parameterization \cite{lw02}. The initial transverse momentum
distribution $g_A(k_T,Q^2,b)$ is assumed to have a Gaussian form
with a width that includes both an intrinsic part in a nucleon and 
nuclear broadening. This parton model can describe high-$p_T$ hadron 
spectra in $p+p(\bar{p})$ well, as shown in Fig.~\ref{fig-ua1th} 
that compares the parton model calculation \cite{Wang:1998ww} with 
the experimental data on the inclusive charged hadron spectra at 
different collider energies.

\begin{figure}
\centerline{\psfig{figure=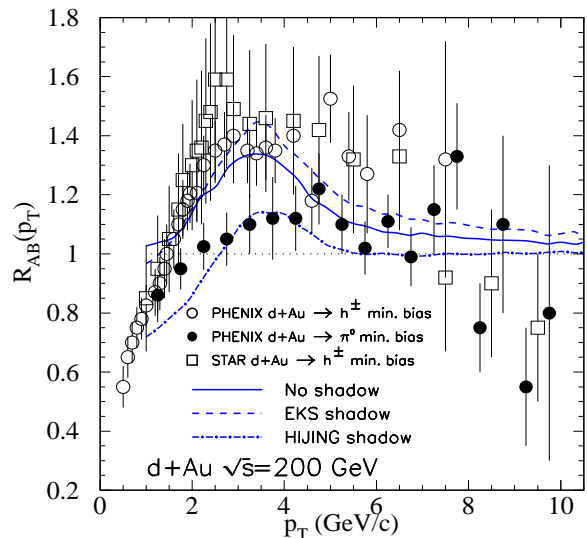,width=3.0in,height=2.8in}}
\baselineskip=10pt
\caption{The first prediction \protect\cite{Wang:1998ww} of
the Cronin effect in $p+Au$ collisions 
at $\sqrt{s}=200$ GeV compared to the recent RHIC data from 
PHENIX \protect\cite{dauphenix} and STAR \protect\cite{daustar}.
The dashed and dot-dashed lines correspond to 
HIJING \protect\cite{lw02} and EKS \protect\cite{eks} 
parameterizations of nuclear parton distributions.
\label{fig:dau}}
\end{figure}

In $p+A$ collisions, this parton model incorporates both nuclear
modification of the parton distributions and broadening of the
initial transverse momentum. The initial momentum broadening
leads to an enhancement, known as the Cronin effect, of the
hadron spectra in the intermediate $p_T$ region. The enhancement
disappears at large $p_T$. The parameters have been fitted to the 
nuclear modification of the $p_T$ spectra in $p+A$ collisions at up
to the Fermilab energy $\sqrt{s}=40$ GeV. Shown in Fig.~\ref{fig:dau}
is the first prediction made in 1998 \cite{Wang:1998ww}
of the Cronin effect at RHIC for $p+Au$ collisions 
at $\sqrt{s}=200$ GeV as compared to the RHIC data on $d+Au$
collisions. As one can see, the initial multiple
scattering in nuclei can give some moderate Cronin enhancement
of the high-$p_T$ spectra, though the details depend on the
parameterization of the nuclear modification of parton distributions.
Therefore, any observed suppression of the high-$p_T$ spectra 
in $Au+Au$ collisions has to be caused by jet quenching.

In $A+A$ collisions, one has to consider medium modification
of the fragmentation functions due to parton energy loss
according to Eq.~(\ref{modfrag}).
Assuming a 1-dimensional expanding medium with a gluon 
density $\rho_g(\tau,r)$ that is proportional to the 
transverse profile of participant nucleons, one can calculate 
the impact-parameter dependence of the energy loss,
\begin{equation}
\Delta E(b,r,\phi)\approx 
\langle \frac{dE}{dL}\rangle_{1d}
\int_{\tau_0}^{\Delta L} d\tau\frac{\tau-\tau_0}{\tau_0\rho_0}
\rho_g(\tau,b,\vec{r}+\vec{n}\tau),
\label{eq-path}
\end{equation}
according to Eq.~(\ref{effloss}), where $\Delta L(b,\vec{r},\phi)$ 
is the distance a jet, produced at $\vec{r}$, has to travel 
along $\vec{n}$ at an azimuthal angle $\phi$ relative to the 
reaction plane in a collision 
with impact-parameter $b$. Here, $\rho_0$ is the averaged 
initial gluon density at initial time $\tau_0$ 
in a central collision and $\langle dE/dL\rangle_{1d}$ 
is the average parton energy loss over a distance $R_A$
in a 1-d expanding medium with an initial uniform gluon 
density $\rho_0$. The corresponding energy loss 
in a static medium with a uniform gluon density 
$\rho_0$ over a distance $R_A$ is \cite{ww02}
$dE_0/dL=(R_A/2\tau_0)\langle dE/dL\rangle_{1d}$.
We will use the parameterization in Eq.~(\ref{eq:loss})
for the effective energy dependence of the
quark energy loss.

\begin{figure}
\centerline{\psfig{figure=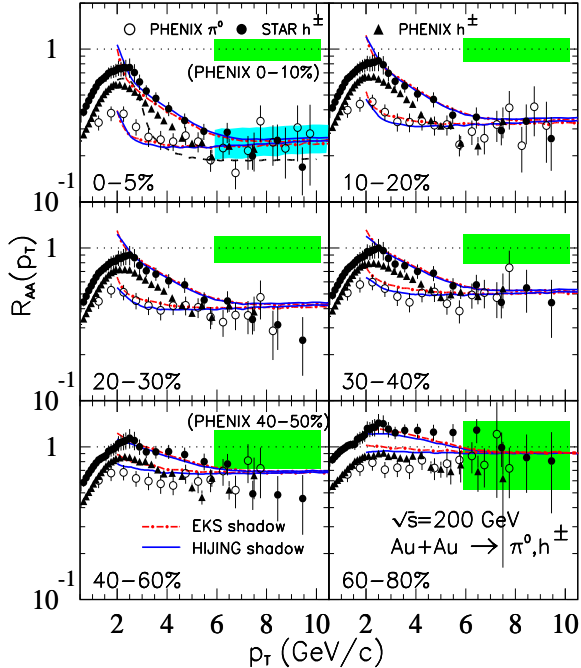,width=3.0in,height=3.5in}}
%\centerline{\includegraphics[angle=-90,width=8.4cm]{r_aacern.eps}
\caption{Hadron suppression factors in $Au+Au$ collisions
as compared to data from STAR\protect\cite{star-r2} and 
PHENIX \protect\cite{phenix-r2}. See text for a detailed explanation.
\label{fig-raa}}
\end{figure}

Shown in Fig.~\ref{fig-raa} are the calculated nuclear 
modification factors
$R_{AB}(p_T)=d\sigma^h_{AB}/\langle N_{\rm binary}\rangle d\sigma^h_{pp}$
for hadron spectra ($|y|<0.5$) in $Au+Au$ collisions 
at $\sqrt{s}=200$ GeV, as compared to experimental 
data \cite{star-r2,phenix-r2}. Here,
$\langle N_{\rm binary}\rangle=\int d^2bd^2r t_A(r)t_A(|\vec{b}-\vec{r}|)$.
To fit the observed $\pi^0$ suppression (solid lines) in the most 
central collisions, we have used $\mu=1.5$ GeV,
$\epsilon_0=1.07$ GeV/fm and $\lambda_0=1/(\sigma\rho_0)=0.3$ fm
in Eqs.~(\ref{modfrag}) and (\ref{eq:loss}).
The hatched area (also in other figures in this report) indicates 
a variation of $\epsilon_0=\pm 0.3$ GeV/fm.
The hatched boxes around $R_{AB}=1$ represent experimental
errors of STAR data in overall normalization, including a constant
factor of about 18\% from the determination of total $p+p$ inelastic
cross section.
Nuclear $k_T$ broadening and parton shadowing together give a slight 
enhancement of hadron spectra at intermediate $p_T=2-4$ GeV/$c$ 
without parton energy loss.

\begin{figure}
\centerline{\psfig{figure=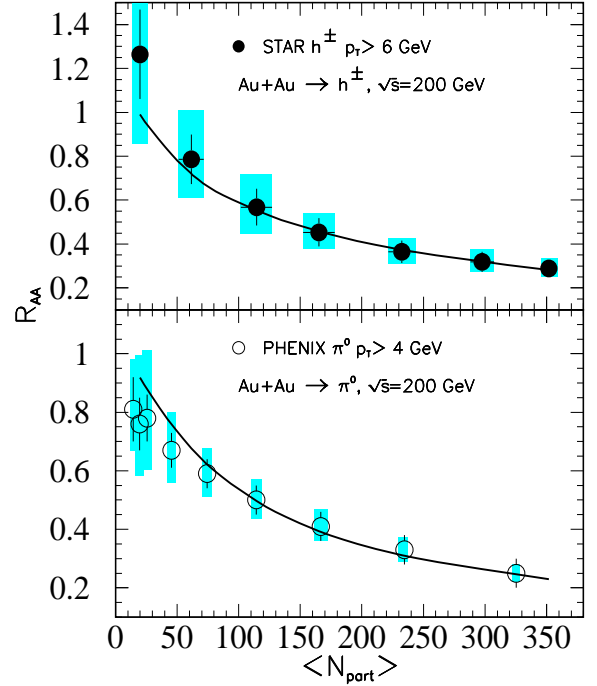,width=3.0in,height=3.6in}}
\caption{The centrality dependence of the measured single inclusive 
hadron suppression \protect\cite{phenix-r2,star-r2} at high-$p_T$ as 
compared to theoretical calculation with parton energy loss}
\label{fig-r-centrality}
\end{figure}

The flat $p_T$ dependence of the $\pi^0$ suppression is 
a consequence of the strong energy dependence of the
parton energy loss as also noted in 
Refs.~\cite{Jeon:2002dv,Muller:2002fa}. 
This is in sharp contrast to the energy
dependence of the suppression in DIS as shown 
in Fig.~\ref{fig:hermes2} and should be a strong indication
of the detailed balance effect in a thermal medium. Such
an effect also causes the slight rise of $R_{AB}$
at $p_T<4$ GeV/$c$ in the calculation. In this
region, one expects the fragmentation picture to gradually 
lose its validity and to be taken over by other non-perturbative 
effects, especially for kaons and baryons.
As a consequence, the $(K+p)/\pi$ ratio in central $Au+Au$
collisions is significantly larger than in peripheral $Au+Au$ or
$p+p$ collisions \cite{phenix-fl}. This has been attributed 
to parton coalescence in the fragmentation of the leading
parton \cite{Hwa:2002tu,Fries:2003vb,Greco:2003xt}. The slight
flavor dependence of the Cronin effect in $d+Au$ collisions
as shown in Fig.~\ref{fig:dau} could also be attributed to
the coalescence effect \cite{Zhang:2003jr,Hwa:2004zd}.
Such an effect, while providing interesting insight into the 
hadronization mechanism in the quark gluon plasma, will 
complicate the picture of jet quenching 
and introduce large uncertainties in the physics extracted from 
jet tomography analysis of the experimental data. Fortunately, 
the effect has been shown to disappear at large $p_T>5$ GeV/$c$
both in the coalescence models and the experimental data.
The suppression ratio of charged hadrons and $\pi^0$ and of $\Lambda$ 
and $K$ all converge \cite{star-fl} at large $p_T$. 
In the calculation shown in Fig.~\ref{fig-raa}, a nuclear-dependent 
(proportional to $\langle N_{\rm binary}\rangle$) soft
component was added to kaon and baryon fragmentation functions 
to take into account the coalescence effect, so that
$(K+p)/\pi\approx 2$ at $p_T\sim 3$ GeV/$c$ in the most 
central $Au+Au$ collisions and it approaches its $p+p$ value 
at $p_T>5$ GeV/$c$. To demonstrate the sensitivity to the
parameterized parton energy loss in the intermediate $p_T$ region, 
we also show $R_{AA}^{h^{\pm}}$ in 0-5\% centrality (dashed line)
for $\mu=2.0$ GeV and $\epsilon_0=2.04$ GeV/fm without the 
soft component.

The measured centrality dependence of the single hadron
suppression in $Au+Au$ collisions, shown in Fig.~\ref{fig-r-centrality},
agrees very well with parton model with energy loss. This is 
the consequence of the centrality dependence of the
averaged total energy loss in Eq.~(\ref{effloss}),
which leads to an effective surface emission of the surviving 
jets. Jets produced around the core of the overlapped region 
are strongly suppressed, since they lose the largest amount 
of energy. The centrality dependence of the suppression is
found to be more dominated by the geometry of the produced
dense matter than the length dependence of the parton
energy loss.

\begin{figure}
\centerline{\psfig{figure=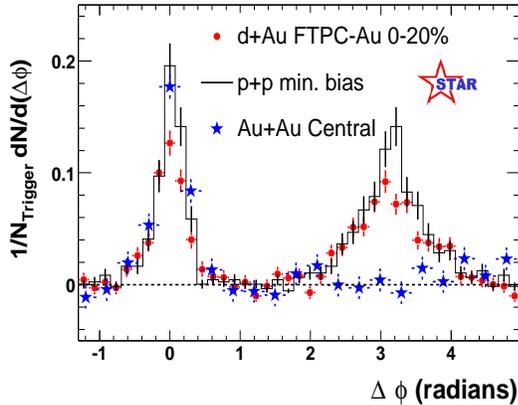,width=3.0in,height=2.2in}}
\caption{ (a) Two-particle azimuthal distributions for minimum bias 
and central $d+Au$ collisions, and for $p+p$ 
collisions\protect\cite{daustar}.
(b) Comparison of two-particle azimuthal distributions for central
d+Au collisions to those seen in p+p and central Au+Au collisions.}
\label{fig-star-phicorr}
\end{figure}

\subsection{Di-hadron Spectra}

Since jets are always produced in pairs in LO pQCD, two-hadron 
correlations in $p+p$ collisions should have unique characteristics 
pertaining to the back-to-back jet structure of the initial hard 
parton-parton scattering. Jet quenching should also modify this 
di-hadron correlation of the back-to-back jets.
Shown in Fig.~\ref{fig-star-phicorr} are the two-hadron correlations in 
azimuthal angle as measured by STAR experiment in $p+p$, $d+Au$ 
and $Au+Au$ collisions at RHIC \cite{daustar}. While the
same-side correlation remains the same, the away-side correlation due 
to the back-side jet is strongly suppressed in central $Au+Au$ collisions.
The energy loss that causes the suppression of the single inclusive
hadron spectra should also be able to explain the disappearance of
away-side correlation.

\begin{figure}
\centerline{\psfig{figure=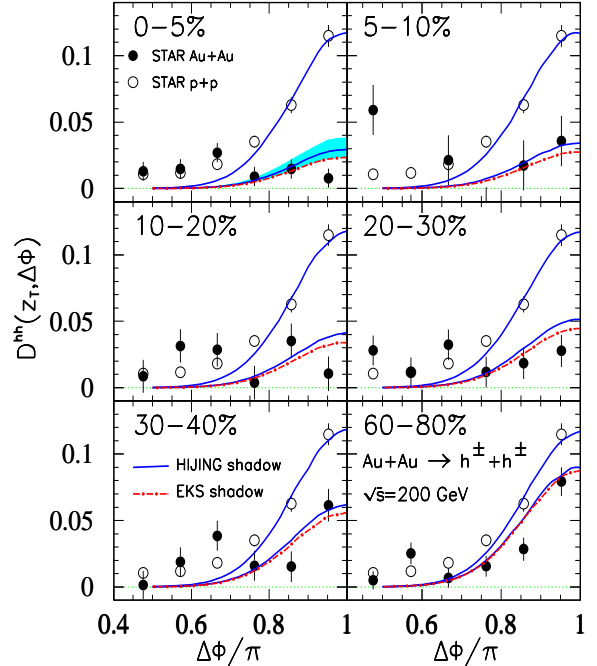,width=3.0in,height=3.5in}}
%\centerline{\includegraphics[angle=-90,width=8.4cm]{phicern.eps}
\caption{Back-to-back correlations for charged hadrons 
with $p^{\rm trig}_T>p_T>2$ GeV/$c$, 
$p^{\rm trig}_T=4-6$ GeV/$c$ and $|y|<0.7$ in $Au+Au$ (lower curves) 
and $p+p$ (upper curves)
collisions as compared to the STAR\protect\cite{star-jet} data.
\label{fig-b2b}}
\end{figure}

In the same LO pQCD parton model, one can calculate the 
spectra \cite{Wang:2003mm},
\begin{eqnarray}
  &&E_1E_2\frac{d\sigma^{h_1h_2}_{AA}}{d^3p_1d^3p_2}=\frac{K}{2}\sum_{abcd} 
  \int d^2b d^2r dx_a dx_b d^2k_{aT} d^2k_{bT}  \nonumber \\
 &\times& t_A(r)t_A(|{\bf b}-{\bf r}|) 
  g_A(k_{aT},r)  g_A(k_{bT},|{\bf b}-{\bf r}|) \nonumber \\
   &\times& f_{a/A}(x_a,Q^2,r)  f_{b/A}(x_b,Q^2,|{\bf b}-{\bf r}|)  
   \nonumber \\
  &\times& D_{h/c}(z_c,Q^2,\Delta E_c)
  D_{h/d}(z_d,Q^2,\Delta E_d) 
 \frac{\hat{s}}{2\pi z_c^2 z_d^2} \nonumber \\
&\times& \frac{d\sigma}{d\hat{t}}(ab\rightarrow cd)
 \delta^4(p_a+p_b-p_c-p_d),
 \label{eq:dih}
\end{eqnarray}
of two back-to-back hadrons from independent fragmentation
of the back-to-back jets. 
Let us assume hadron $h_1$ is a triggered hadron 
with $p_{T1}=p_T^{\rm trig}$. One can define a hadron-triggered 
fragmentation function (FF) as the back-to-back correlation with 
respect to the triggered hadron:
\begin{equation}
  D^{h_1h_2}(z_T,\phi,p^{\rm trig}_T)=
  p^{\rm trig}_T \frac{d\sigma^{h_1h_2}_{AA}/d^2p^{\rm trig}_T dp_Td\phi}
  {d\sigma^{h_1}_{AA}/d^2p^{\rm trig}_T},
\end{equation}
similarly to the direct-photon triggered FF \cite{wh} 
in $\gamma$-jet events. Here, $z_T=p_T/p^{\rm trig}_T$ and 
integration over $|y_{1,2}|<\Delta y$ is implied. 
In a simple parton model, two jets should be
exactly back-to-back. The initial parton transverse momentum distribution
in our model will give rise to a Gaussian-like angular distribution.
In addition, we also take into account transverse momentum smearing
within a jet using a Gaussian distribution with a width of
$\langle k_\perp\rangle=0.6$ GeV/$c$.

Shown in Fig.~\ref{fig-b2b} are the calculated back-to-back correlations 
for charged hadrons in $Au+Au$ collisions as compared to STAR 
data \cite{star-jet}. The same energy loss that is used to calculate 
single hadron suppression can also describe
well the observed away-side hadron suppression and its centrality
dependence.

% In the data, a background 
%$B(p_T)[1+2v_2^2(p_T)\cos(2\Delta\phi)]$ from uncorrelated hadrons
%and azimuthal anisotropy has been subtracted. The value of $v_2(p_T)$
%is measured independently while
%$B(p_T)$ is determined by fitting the observed correlation in the
%region $0.75<|\phi|<2.24$ rad \cite{star-jet}.

With cross sections integrated over $\phi$, one obtains a 
hadron-triggered FF $D^{h_1h_2}(z_T,p^{\rm trig}_T)$. 
The suppression factor 
$I_{AA}(z_T, p_T^{\rm trig})\equiv D^{h_1h_2}_{AA}/D^{h_1h_2}_{pp}$ 
defined by the STAR experiment \cite{star-jet}
is just the medium modification factor of the hadron-triggered FF.
The shape of the modification is predicted \cite{Wang:2003mm}
to be very similar to that of the $\gamma$-triggered fragmentation
functions \cite{wh}.

\subsection{High $p_T$ Azimuthal Anisotropy}

Another aspect of jet quenching is the azimuthal anisotropy
of the spectra caused by parton energy loss which depends on the
azimuthal angle according to Eq.~(\ref{eq-path}).
In non-central collisions, the average path length of parton
propagation will vary with the azimuthal angle relative to
the reaction plane. This leads to an azimuthal
dependence of the total parton energy loss and therefore
azimuthal asymmetry of high-$p_T$ hadron spectra \cite{wangv2,gvwv2}.
Such an asymmetry is another consequence of parton energy loss
and yet it is not sensitive to effects of initial state interactions. 

\begin{figure}
\centerline{\psfig{figure=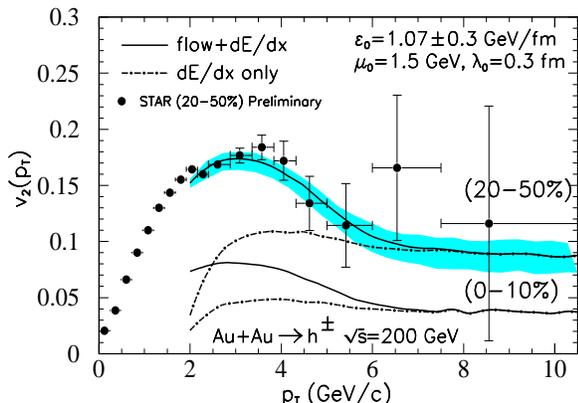,width=3.0in,height=2.1in}}
\caption{Azimuthal anisotropy in $Au+Au$ collisions
as compared to the STAR\protect\cite{Snellings:2003mh} 4-particle 
cumulant result.}
\label{fig-highptv2}
\end{figure}

In the parton model, the azimuthal angle dependence of the parton
energy loss will be given by Eq.~(\ref{effloss}) for non-central
heavy-ion collisions. Using this anisotropic energy loss in
the effective modified fragmentation functions in the pQCD
parton model of high-$p_T$ hadron spectra, one will obtain
azimuthal anisotropic spectra.
Shown in Fig.~\ref{fig-highptv2} is $v_2(p_T)$ 
(second Fourier coefficient of the azimuthal angle distribution) 
of charged hadrons generated from
parton energy loss (dot-dashed) as compared to preliminary 
STAR data \cite{Snellings:2003mh},
using the 4-particle cumulant moments method \cite{Borghini:2001vi} which
is supposed to reduce non-geometrical effects such as
inherent two-particle correlations from di-jet 
production \cite{Kovchegov:2002nf}.
The energy loss extracted from high-$p_T$ hadron spectra 
suppression can also account for the observed azimuthal anisotropy 
at large $p_T$. At intermediate $p_T$, the observed $v_2$ is
larger than the simple parton model calculation. Such discrepancy 
can also be attributed to effects of parton coalescence. 
If the difference is made up by kaons and baryons from the
coalescence contribution, one finds that they must 
have $v_2\approx 0.23$ (0.11) for 20-50\% (0-10\%)
collisions. The total $v_2(p_T)$ is shown by the solid lines.

\section{Jet Quenching and QGP}

From both single and di-hadron spectra and their azimuthal
anisotropy $v_2(p_T)$, the extracted average energy loss in the
parton model calculation for a 10 GeV quark in the expanding medium is 
$\langle dE/dL\rangle_{1d}\approx 0.85 \pm  0.24$ GeV/fm, which
is equivalent to $dE_0/dL\approx 13.8 \pm 3.9$ GeV/fm in a static and
uniform medium over a distance $R_A=6.5$ fm at an initial 
time $\tau_0=0.2$ fm.  Compared to the energy loss extracted from 
HERMES data on DIS, this value is about 30 times higher than the 
quark energy loss in cold nuclei. Since the parton energy loss in the 
thin plasma limit is proportional to the gluon number density, one can
conclude that the initial gluon density reached in the central
$Au+Au$ collisions at 200 GeV should be about 30 times higher than
the gluon density in a cold $Au$ nucleus. This number is consistent
with the estimate from the measured rapidity density of charged 
hadrons \cite{Back:2001ae} using the Bjorken scenario \cite{Bjorken:1982qr}
and assuming duality between the number of initial gluons and
final charged hadrons. Given the measured total transverse
energy $dE_T/d\eta\approx 540$ GeV or about 0.8 GeV per charged
hadron \cite{Adcox:2001ry} in central $Au+Au$ collisiona 
at $\sqrt{s}=130$ GeV, the initial energy density in the
formed dense matter is at least 100 times higher than in
cold nuclear matter.

The above analyses of RHIC data on jet quenching are
based on an assumption that the dominant mechanism of the jet 
quenching is parton energy loss before hadronization.
It is reasonable to ask whether leading hadrons 
from the jet fragmentation could have strong interaction with 
the medium and whether hadron absorption could be the main cause 
for the observed jet quenching \cite{Falter:2002jc,Gallmeister:2002us}. 
A detailed analysis of jet quenching data concludes \cite{Wang:2003aw} 
that the data are not compatible with such a scenario of hadronic 
absorption and that the observed patterns of jet quenching in heavy-ion 
collisions at RHIC are the consequences of parton energy loss.

In addition to the long hadron formation time (it could be about 
30-70 fm/$c$ for a 10 GeV pion) estimated from uncertainty principle, 
there are many patterns of the suppression that can rule out the
scenario of hadron absorption. Foremost among them, the 
large $v_2$ at high-$p_T$ and the same-side di-hadron correlation 
cannot be reconciled with the hadron absorption scenario. 

Since the spectra azimuthal anisotropy is caused by the 
eccentricity of the dense medium, which decreases 
rapidly with time \cite{Kolb:2000sd} due to transverse expansion, 
it has to happen at a  very early time. By the time of a few fm/$c$ 
before the formation of any hadron, the eccentricity is already
reduced to a non-significant value. Any further interaction cannot
produce much spectral anisotropy. 

\begin{figure}
\centerline{\psfig{figure=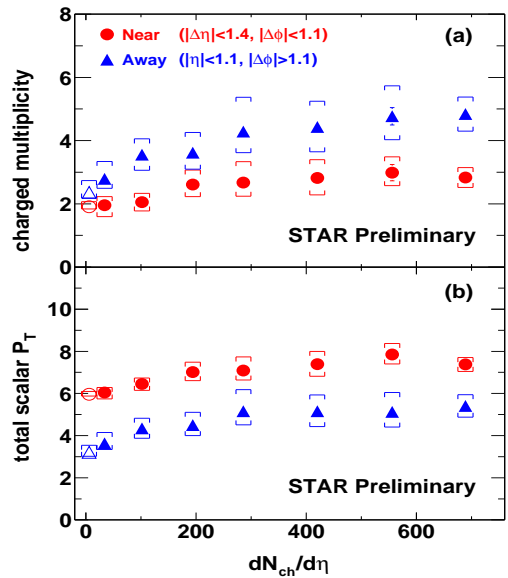,width=3.0in,height=3.1in}}
\caption{Charged hadron multiplicity (a) and total scalar $p_T$ (b)
of the near and away side jet with $4<p_T^{trig}<6$ GeV/$c$ 
and $0.15<p_T<4$ GeV/$c$ in $p+p$ (open symbols) and $Au+Au$
collisions from STAR experiment\protect\cite{fqwang}.}
\label{fqwang}
\end{figure}

As shown in Fig.~\ref{fig-star-phicorr}, while the away-side jet
is suppressed in central $Au+Au$ collisions, the same-side di-hadron 
correlation remains almost the same as in $p+p$ and $d+Au$ collisions.
This is clear evidence that jet hadronization takes place 
outside the dense medium with a reduced parton energy. A recent
study \cite{majumder} of the di-hadron fragmentation functions 
shows that the conditional (or triggered) di-hadron distribution at
large $z$ within a jet is quite stable against radiative
evolution (or energy loss). On the other hand,
a hadron absorption mechanism will suppress both the leading
and secondary hadron in the same way and should lead to the
suppression of the same-side correlation. One may argue that
the jet whose leading hadron is the trigger hadron could
be emitted from the surface and suffer no energy loss
and therefore no suppression of the secondary hadron. This
is not true. Even though the same-side jet is produced close
to the surface due to trigger bias, the parton jet on the average 
still loses about 2 GeV energy \cite{Wang:2003aw}, which
should be carried by soft hadrons in the direction of the 
triggered hadron. If one collects all the energy from jet
fragmentation for a fixed value of trigger hadron $p_T$,
this energy in central $Au+Au$ collisions should be larger
than in $p+p$ collisions by about 2 GeV. This has indeed 
been confirmed by preliminary data from STAR \cite{fqwang}, as 
shown in Fig.~\ref{fqwang}, where the scalar sum of the $p_T$
on the same side of the triggered hadron is indeed about 2 GeV
larger in central $Au+Au$ than in $p+p$ collisions. The
analysis \cite{fqwang} also sees modification of the fragmentation
function both along and opposite the trigger hadron direction.

Finally, if hadron absorption can suppress high-$p_T$
hadrons and jets, it would most likely happen in heavy-ion
collisions at the SPS energy. Hadron spectra at this energy 
are very steep at high-$p_T$ and are very sensitive to initial 
transverse momentum broadening and parton energy 
loss \cite{Wang:1998ww}. However, the measured $\pi$ spectrum in 
central $Pb+Pb$ collisions only shows the expected Cronin 
enhancement \cite{wa98,wang98h} with no sign of significant 
suppression. More recent analysis of the $Pb+Pb$
data at the SPS energy also shows \cite{ceres} that both same-side 
and back-side jet-like correlations are not suppressed, though the
back-side distribution is broadened.

\section{Summary and Outlook}

In summary, with the latest measurements of high-$p_T$ hadron spectra and
jet correlations in $p+p$, $d+Au$ and $Au+Au$ collisions at RHIC, the 
observed jet quenching in $Au+Au$ collisions has been established 
as a consequence of final-state interaction between jets and the 
produced dense medium. The collective body of data, suppression of 
high-$p_T$ spectra and back-to-back jet correlation, high-$p_T$ 
anisotropy and centrality dependence of the observables, points 
to parton energy loss as the culprit of the observed  jet quenching.
A simultaneous phenomenological study within a LO pQCD
parton model incorporating the parton energy loss describes 
the experimental data of $Au+Au$ collisions very well. The
extracted average energy loss for a 10 GeV quark in the expanding 
medium is equivalent to an energy loss in a static and uniform 
medium that is about 30 times larger than in a cold nucleus. This
leads us to conclude that the initial gluon (energy) density
is about 30 (100) times of that in a cold nuclear matter.
It is inconceivable to imagine within our current understanding 
of QCD that any form of matter  other than quark gluon plasma 
could exist at such a high energy density.

A vast collection of data on bulk properties of particle
production in heavy-ion collisions provides further evidence
of QGP formation at RHIC. Most striking among these data is the
elliptic flow measurement \cite{star-flow1} that was found to
saturate the hydrodynamic limit and exhibit all the
signature behaviors of a collective hydro flow, including
the splitting in $v_2(p_T)$ of hadrons with 
different masses \cite{phenix-fl2,star-fl,Sorensen:2003kp}. 
Together with jet quenching, they provide a collection of 
compelling evidence for the formation of a strongly interacting 
quark gluon plasma at RHIC.

\begin{figure}
\centerline{\psfig{figure=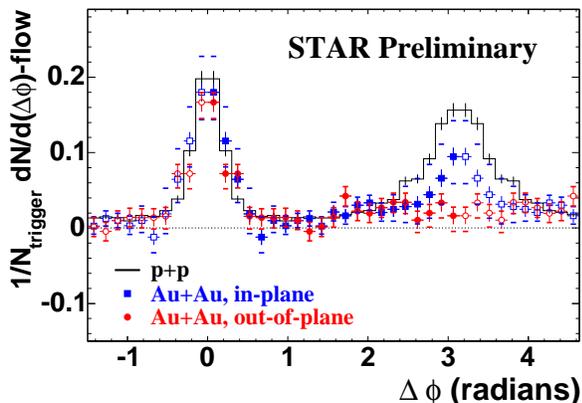,width=3.0in,height=2.1in}}
\caption{High-$p_T$ two-hadron correlation in azimuthal angle for
different orientation of the trigger hadron with respect to the
reaction \protect\cite{filimonov}.}
\label{fig-star-inout}
\end{figure}

The discovery of jet quenching and its central role in 
characterizing the strongly interacting quark gluon plasma 
will usher in another new era of studying the properties of the 
dense matter. With higher-precision data extending to higher $p_T$,
one can find out the missing energy of the suppressed jet \cite{pratt}
and map out in detail the modification of the fragmentation
functions and the energy dependence of the energy loss.
The preliminary results from STAR \cite{fqwang} 
on modified fragmentation functions in $Au+Au$ collisions
have already demonstrated the potential of the study and
could even shed light on parton thermalization and hadronization.
The ultimate measurement of a modified FF in heavy-ion collisions 
will be the direct-$\gamma$-triggered FF. Recent preliminary
results from PHENIX \cite{frantz} indicate that such a measurement
will be within reach in the next few years.

One can also study the di-hadron correlations with
respect to the reaction plane. Shown in Fig.~\ref{fig-star-inout}
are the preliminary results from STAR \cite{filimonov} 
on di-hadron correlation in azimuthal angle for different 
orientations of the trigger hadron with respect to the reaction 
plane in $Au+Au$ collisions. This could eventually measure the length
dependence of the parton energy loss. Measurements of charmed 
mesons can provide additional tests of the
parton energy loss scenario of jet quenching, since recent theoretical
studies \cite{Dokshitzer:2001zm,Djordjevic:2003qk,Zhang:2003wk,Armesto:2003jh} show many unique features of heavy quark energy loss that are different 
from those of light quarks and gluons.

\vspace{-0.3in}

\section*{Acknowledgement}

\vspace{-0.2in}

This work is supported  by
the Director, Office of Energy
Research, Office of High Energy and Nuclear Physics, Divisions of 
Nuclear Physics, of the U.S. Department of Energy under Contract No.
DE-AC03-76SF00098.

\end{multicols}

\end{document}